\documentclass[final]{aipproc}
\layoutstyle{6x9}

\usepackage{amssymb,amsmath}

\begin{document}

\title{Neutrino-induced coherent pion production off nuclei}



\author{T.~Leitner}{address={Institut f\"ur Theoretische Physik, Universit\"at Giessen,
    Germany}}

\author{U.~Mosel}{address={Institut f\"ur Theoretische Physik, Universit\"at Giessen,
    Germany}}

\author{S.~Winkelmann}{address={Institut f\"ur Theoretische Physik, Universit\"at Giessen,
    Germany}}

\begin{abstract}
  All available theoretical estimates of neutrino-induced coherent pion production rely on the
  'local approximation' for the Delta propagator.  The validity of this approximation is
  scrutinized. It is found that the local approximation overestimates the
  neutrino-induced coherent pion production on nuclei significantly, by up to 100\%.
\end{abstract}

\maketitle

\section{Introduction}

By scattering electroweak probes with nuclei, pions can be produced either coherently,
leaving the nucleus intact, or incoherently. The former one has attracted considerable
attention in the last years, both theoretically
\cite{Singh:2006bm,AlvarezRuso:2007tt,Amaro:2008hd,Berger:2008xs,Paschos:2009ag} and
experimentally \cite{Hasegawa:2005td,Hiraide:2008eu,AguilarArevalo:2008xs}. While there is
compelling evidence for NC coherent pion production, no evidence for CC coherent pion
production could be found. However, all these experimental analyses suffer from the fact
that the coherent fraction is not accessible directly but has to be extracted from data
assuming specific models for incoherent pion production. Furthermore, the theoretical
models for coherent scattering used in the experimental analyses overpredict the measured
rates.

The above mentioned theoretical models can be classified into two classes: first, the PCAC
models which relate the coherent pion production to a forward scattering amplitude via
PCAC assuming that specific nuclear effects play no role, besides providing nuclear size
information. Second, models based on nuclear structure which start from a theoretical
description of the nuclear structure and sum the pion production amplitude coherently over
all target nucleon states. Both classes rely on the so-called local approximation which
allows one to factorize out the nuclear form factor. In the following we investigate the
impact of the local approximation for neutrino-induced processes. For further details, we
refer the reader to \cite{Leitner:2009ph} and references therein.

\section{Full calculation vs. local approximation}

Our model assumes that pions are dominantly created via the $\Delta(1232)$ resonance.
Then, the hadronic current for a nucleon is given by
\begin{equation}
\label{Jhadr}
J^\mu_{\rm nucleon}(p,q) = i  \frac{f^*}{m_\pi} C^\Delta F(p_\Delta^2)\,\bar{u}(\vec{p}\,') k_\pi^\alpha G_{\alpha \beta}(p_\Delta) \Gamma^{\beta \mu}(p,q) u(\vec{p}),
\end{equation}
with the pion momentum $k_\pi$, the nucleon's final and initial momenta, $p'$ and $p$, and
the transferred four-momentum $q=p'-p$. Thus, the $\Delta$ momentum reads $p_\Delta = p +
q$. $G_{\alpha \beta}$ represents the full Rarita-Schwinger propagator
\begin{equation} \label{Deltaprop} G_{\alpha \beta} = \frac{1}{p_\Delta^2 - M_\Delta^2 + i
    M_\Delta \Gamma_\Delta} \, P_{\alpha \beta} ~,
\end{equation}
where $P_{\alpha \beta}$ is the usual Rarita-Schwinger projection operator.  The vertex
function $\Gamma^{\beta \mu}$ denotes the standard electroweak vertex structure with
vector and axial contributions including the resonance excitation form factors.
$f^*/m_\pi$ is the $N \Delta \pi$ coupling constant, $F(p_\Delta^2)$ a form factor for the
$\Delta$ and $C^\Delta$ contains isospin factors (cf.\ \cite{Leitner:2009ph} for details).

The single particle current (\ref{Jhadr}) has to be summed over all occupied
single-particle states in the target nucleus (full calculation), so that
\begin{align} \label{Jnucl}
J^\mu_{\rm nucleus}(q) &= \sum_{i} \int d^3\!p \, J^\mu_{i}(p,q) \nonumber \\
&= i  \frac{f^*}{m_\pi} C^\Delta \sum_{i} \int d^3\!p \,F(p_\Delta^2)\, \bar{\psi_i}(\vec{p}\,') k_\pi^\alpha G_{\alpha
    \beta}(p_\Delta) \Gamma^{\beta \mu}(p,q) \psi_i(\vec{p}),
\end{align}
where the bound-state-spinors $\psi_i(\vec{p})$ are obtained in a Walecka-type mean field
model \cite{Peters:1998mb} and replace the free-particle-spinors $u(\vec{p})$ in
(\ref{Jhadr}) (same for $\bar{u}(\vec{p}\,')$). Note that the momentum integration extends
also over the $\Delta$ propagator since $p_\Delta = p + q$.

The 'local approximation' now consists of fixing the momentum of the initial nucleon state
in the product $G_{\alpha \beta}(p_\Delta) \Gamma^{\beta \mu}(p,q)$ to some value --- here
we use
\begin{equation}
 \vec{p}\,^0 = -(\vec{q} - \vec{k}_\pi)/2 \quad \Rightarrow \quad \vec{p}\,^0_\Delta = (\vec{q} + \vec{k}_\pi)/2.
\end{equation}
With that, the momentum of the $\Delta$ resonance is determined, and the $\Delta$
propagator can be moved out of the integral and even out of the sum. This approximation
basically consists of suppressing the propagation of the $\Delta$ resonance and
corresponds to the assumption of a very heavy $\Delta$ resonance. Consequently, the $W,Z +
N \to \pi + N$ vertex becomes local.  In an $r$-space representation, the current in the
local approximation reads
\begin{equation} \label{Jlocal}
\widetilde{J}\;^\mu_{\rm nucleus}(q) =  i  \frac{f^*}{m_\pi} C^\Delta \frac{k_\pi^\alpha}{{p^0_\Delta}^2 - M_\Delta^2 + i
    M_\Delta \Gamma_\Delta} \,
  \int d^3\!r \, e^{i(\vec{q} - \vec{k}_\pi) \cdot \vec{r}}\,
  {\rm tr} \left( \rho(\vec{r},\vec{r}) P_{\alpha \beta}(p^0_\Delta) \Gamma^{\beta
      \mu}(p^0,q)\right).
\end{equation}
Here the trace is taken over the Dirac indices and $\rho(\vec{r},\vec{r})$ is the diagonal
element of the one-body density matrix. This is the final result in the local
approximation.  Equation (\ref{Jlocal}) shows that the nuclear form factor has been
factorized out; all the other (non-local) densities present in the full expression
(\ref{Jnucl}) no longer appear.

\section{Results}

In the following, we compare the full calculation, based on Eq.\ (\ref{Jnucl}) with a
propagating $\Delta$, with the results of the local approximation [Eq.\ (\ref{Jlocal})]
for the target nucleus $^{12}$C. To isolate the effects of the local approximation, both
calculations are done in the plane wave approximation in which the produced pion is taken
to be a free particle. We also do not include in-medium changes of the $\Delta$ spectral
function in the propagator [Eq.~(\ref{Deltaprop})]. Both calculations use the same nuclear
structure model, i.e., the density and momentum distributions are calculated consistently
in the same relativistic mean field model. 

\begin{figure}
\includegraphics[width=85mm]{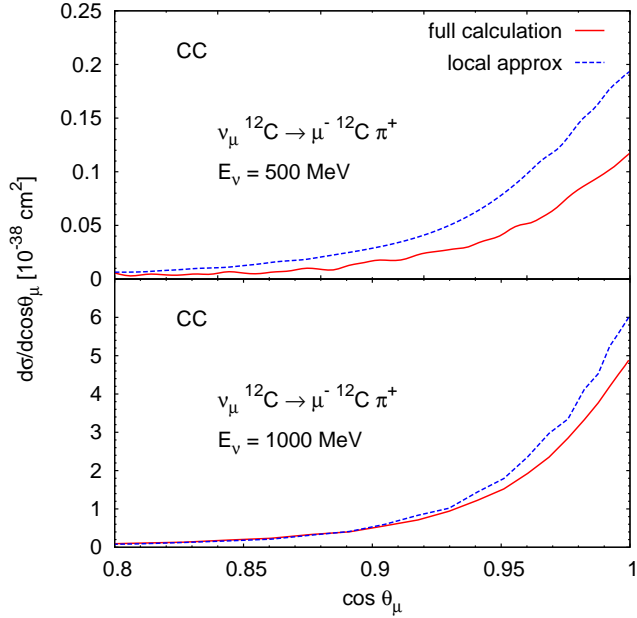}
\caption{\label{fig:dsigmadcos} CC induced pion angular distribution for a
  neutrino energy of 500 MeV (1000 MeV) and target $^{12}$C. The dashed curve gives the result of the
  calculation using the local approximation [cf.\ Eq.\ (\ref{Jlocal})]; the solid curve
  gives the result of a fully dynamic calculation [cf.\ Eq.\ (\ref{Jnucl})]. All curves are
  without pion final state interactions.}
\end{figure}

Figure \ref{fig:dsigmadcos} shows a comparison of the full calculation with the results
obtained by using the local approximation for the angular distribution of the produced
pions at $E_\nu=500$ and 1000 MeV. The difference between the full and the approximate
calculation is larger at lower energies. At $E_\nu=500$ MeV, the difference is dramatic
over a wide angular range and amounts to a factor of $\approx 1.7$ at zero degrees. At
$E_\nu=1000$ MeV, we find that the local approximation gives a cross section for very
forward angles that is about 20\% larger than that obtained in the full calculation.

\begin{figure}
\includegraphics[width=85mm]{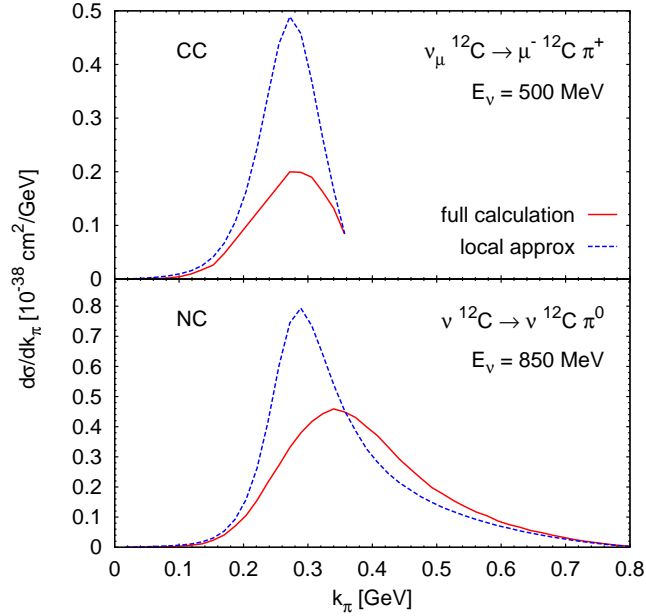}
\caption{\label{fig:dsigmadk} Pion momentum distribution for 500 (CC) and 850 (NC) MeV
  incoming neutrino energy. The solid and the dashed curves are as in the previous
  figure.}
\end{figure}

The pion momentum distribution for induced by CC muon neutrinos of $E_\nu=500$ MeV is
shown in the top panel of Fig.\ \ref{fig:dsigmadk}. It is seen that the local
approximation overestimates the full result by a factor of about 2.5 at the peak.  We find
qualitatively similar results for NC induced coherent pion production (bottom panel). The
slight shift downward relative to the fully dynamical result is a consequence of the local
approximation which assumes a very heavy $\Delta$ thus minimizing any recoil effects. We
finally note that our curves agree quantitatively with the recent results of the Ghent
group \cite{praetphd}.
While all these results were obtained in calculations without pion final state
interactions (fsi) the recent calculations by Nakamura et al.\cite{Nakamura} show that the
local approximation fails as badly when the pion fsi are taken into account.

\section{Conclusions}
The local approximation, used from the start in all presently available microscopic
calculations, overestimates the coherent neutrino-induced pion production significantly
and involves errors which can reach up to 100\% in the neutrino energy regime relevant to
present experiments (MiniBooNE, T2K).
\\ $\,$ \\
This work has been supported by the Deutsche Forschungsgemeinschaft (DFG).

\end{document}